\documentclass[aps,prb,twocolumn,groupedaddress,citeautoscript]{revtex4}
\usepackage{graphicx}
\begin{document}
\title{The Half-Metallicity of Zigzag Graphene Nanoribbons with Asymmetric Edge Terminations}
\author{Zuanyi Li, Bing Huang, and Wenhui Duan\footnote{Author to whom correspondence should
be addressed. E-mail: dwh@phys.tsinghua.edu.cn}}
\affiliation{Department of Physics, Tsinghua University, Beijing
100084, People's Republic of China}
\date{11 May 2009}
\begin{abstract}

\noindent{The spin-polarized electronic structure and
half-metallicity of zigzag graphene nanoribbons (ZGNRs) with
asymmetric edge terminations are investigated by using first
principles calculations. It is found that compared with symmetric
hydrogen-terminated counterparts, such ZGNRs maintain a
spin-polarized ground state with the anti-ferromagnetic
configuration at opposite edges, but their energy bands are no
longer spin degenerate. In particular, the energy gap of one spin
orientation decreases remarkably. Consequently, the ground state of
such ZGNRs is very close to half-metallic state, and thus a smaller
critical electric field is required for the systems to achieve the
half-metallic state. Moreover, two kinds of studied ZGNRs present
massless Dirac-fermion band structure when they behave like
half-metals.}

\


\end{abstract}


\maketitle


\noindent \textbf{1. INTRODUCTION}

\

\noindent Due to recent progress in preparing single graphite
layer\cite{Novo-S,Berger,Kim-EFE,Novo-N,Kim-N}, graphene, this
two-dimension (2D) electronic system has attracted extensive
interest. Importantly, graphene can be patterned via standard
lithographic technique into new carbon based quasi-one-dimension
materials\cite{Berger-S,Kim-gap}, graphene nanoribbons (GNRs), which
have many properties similar to Carbon nanotubes (CNTs), such as
energy gap dependence of widths and crystallographic
orientations\cite{Kim-gap,Barone,Louie-gap}, electronic confinement
and long phase coherence length\cite{Berger-S}. Furthermore,
different from CNTs, the planar geometry of GNRs allows for the more
straightforward fabrication and better control of crystallographic
orientation. All of these characteristics of GNRs provide an
exciting possibility for the flexible design of GNR-based electronic
devices in nanometer scale\cite{Yan,Huang-JPCC}.

As one kind of GNRs, zigzag graphene nanoribbons (ZGNRs) have many
peculiar and interesting properties because of the presence of the
zigzag
edges\cite{Nakada,Fujita99,Kobayashi,Niimi,Yamashiro,Lee,Louie-gap,Louie-N,Yang-hm-APL,Dutta-hm,Hod-hm,Huang-PRL,Jiang}.
On the one hand, ZGNRs have localized edge
states\cite{Nakada,Fujita99,Kobayashi,Niimi} and exhibit a
spin-polarized semiconducting ground
state\cite{Louie-gap,Yamashiro,Lee,Louie-N}, which is characterized
by the anti-ferromagnetic (AF) configuration with opposite spin
orientations between ferromagnetically ordered edge states at each
edge. Particularly, although the energy bands of the ground state
are spin degenerate, under an appropriate in-plane electric field
perpendicular to the direction of the ribbon edge, ZGNRs are forced
into a half-metallic state where one spin exhibits a metallic
behavior, while the other experiences an increase in the
bandgap\cite{Louie-N,Yang-hm-APL,Dutta-hm,Hod-hm}. So ZGNRs could
serve as spin filter devices in future nanospintronics. On the other
hand, the edges of ZGNRs have unique chemical
reactivity\cite{Jiang}, and might be terminated by many chemical
groups such as hydroxyl, carboxyl and epoxide
groups\cite{He,Lerf,Radovic,Stankovich}. Moreover, it is predicted
that edge oxidation could enhance the half-metallicity of
ZGNRs\cite{Hod}, and that asymmetric terminations could
significantly change the electronic, magnetic and transport
properties of
ZGNRs\cite{Lee,Kusakabe,Yang-hm-JACS,Sodi-hm-PRB,Li-PRL}. Therefore,
the zigzag edges and edge termination groups (including their
symmetries) play a significant role in the properties of ZGNRs, and
it is essential to gain better understanding of the effect of
asymmetric chemical edge terminations on the electronic properties
of ZGNRs, especially the half-metallicity.

In this work, we investigate the relative stabilities, electronic
structure and half-metallic nature of ZGNRs with asymmetric edge
terminations, and the termination groups being considered include
Hydrogen (H), hydroxyl (OH) and carboxyl (COOH). It is found that OH
and COOH terminations are as stable as H termination. Compared with
symmetric H-terminated ZGNRs which have been studied
extensively\cite{Louie-gap,Nakada,Fujita99,Lee,Louie-N}, the ZGNRs
we focus on maintain a spin-polarized ground state with the
anti-ferromagnetic configuration, but their energy bands are no
longer spin degenerate. Particularly, the energy gap for one spin
orientation reduces remarkably, which leads to the fact that the
ground state is very close to the half-metallic state. We show that
such feature arises from the difference of electrostatic potentials
at two edges induced by asymmetric edge terminations, so a smaller
critical electric field compared with fully H-terminated ZGNRs is
required to force the system into half-metallic state. These
properties indicate that the half-metallicity of ZGNRs can be
enhanced by breaking the symmetry of edge terminations via chemical
treatments, and offer a possibility for the design of spin filter
devices based on asymmetric terminated ZGNRs.

\

\noindent \textbf{2. CALCULATION METHOD AND MODEL}

\

\noindent Our electronic structure calculations were performed using
the SIESTA method\cite{Soler}, which is based on the {\it ab initio}
density-functional theory (DFT) within the local spin density
approximation (LSDA)\cite{PZ}. The structure optimizations were
first carried out until atomic forces converged to 0.02
eV/$\rm{\AA}$, and the mesh cutoff energy is chosen as 200 Ry. To
simulate an isolated graphene nanoribbon, we adopted a
three-dimensional repeating model in which each nanoribbon is
separated by 8 \AA\ for both edge-edge and layer-layer intervals,
and the supercell length $a$ along $z$ (periodic) direction of ZGNRs
(Fig. 1) is set to be 2.461 \AA . Moreover, a periodic
saw-tooth-type potential along $x$ direction is used to simulate the
external in-plane electric fields in a supercell.

\begin{figure}
\includegraphics[width=0.475\textwidth]{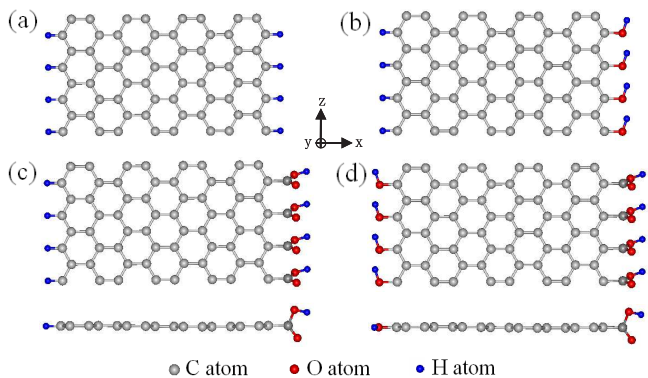}
\caption{Optimized geometry of zigzag graphene nanoribbons with
different edge terminations. (a) 8-ZGNR-H-H. (b) 8-ZGNR-H-OH. (c)
8-ZGNR-H-COOH. (d) 8-ZGNR-OH-COOH. In (c) and (d), both top and side
views are shown. The gray, red, and blue balls represent Carbon,
Oxygen, and Hydrogen atoms respectively. The $z$ axis is taken along
the periodic direction of ZGNRs, and the $x$ axis is along the
direction of external electric field.}\label{structure}
\end{figure}

In accordance with previous
convention\cite{Louie-gap,Nakada,Fujita99,Yamashiro,Lee,Louie-N},
the ZGNRs are classified by the number of zigzag C-C chains forming
the width of the ribbon, and the ZGNR with $n$ zigzag C-C chains is
named as $n$-ZGNR. Herein symmetric H-terminated ZGNRs are named as
ZGNRs-H-H (as shown in Fig. 1(a)), and three kinds of ZGNRs with
asymmetric edge terminations are mainly considered in our study:

\noindent (1) the left and right edges are respectively terminated
by H and hydroxyl group (ZGNRs-H-OH), as shown in Figure 1(b);

\noindent (2) the left and right edges are respectively terminated
by H and carboxyl group (ZGNRs-H-COOH), as shown in Figure 1(c);

\noindent (3) the left and right edges are respectively terminated
by hydroxyl and carboxyl groups (ZGNRs-OH-COOH), as shown in Figure
1(d).

\

\noindent \textbf{3. RESULTS AND DISCUSSION}

\

\noindent We start by studying the relative stability of the ground
states of ZGNRs terminated by different functional groups. First,
for the ZGNRs terminated by OH or COOH groups, the localized edge
states still exist because of the maintenance of $sp^2$
hybridization of edge carbon atoms, and their ground states are also
spin-polarized and in anti-ferromagnetic configuration like fully
H-terminated ZGNRs. In addition, our calculations show that the
binding energies\cite{BE} of a 8-ZGNR-H (one edge is terminated by
H, while the other is bare) and three kinds of functional groups (H,
OH and COOH) are -5.7, -6.3, and -5.5 eV, respectively. These
results indicate that OH and COOH terminations do not change the
spin configuration of the ground state of ZGNRs, and they are as
stable as H termination.

\begin{figure}
\includegraphics[width=0.475\textwidth]{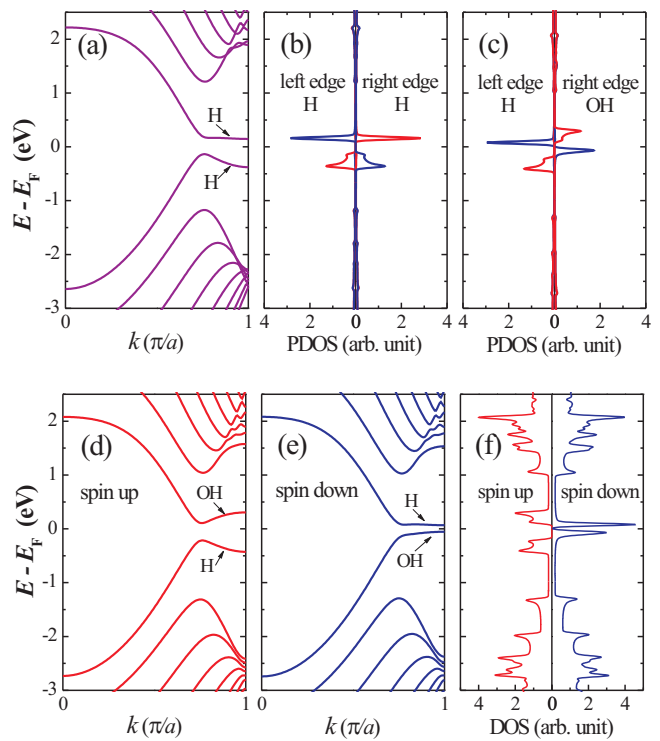}
\caption{(a) The spin-degenerate energy band structure of 8-ZGNR-H-H
in AF configuration. (b), (c) PDOS of two edge carbon atoms in
8-ZGNR-H-H and 8-ZGNR-H-OH respectively, where red (blue) lines are
associated with spin up (down). (d) Spin-up and (e) spin-down energy
bands, and (f) the total DOS of 8-ZGNR-H-OH in AF configuration. In
(a), (d) and (e), the arrows indicate the edge states and which edge
they are localized at.}\label{z8-H-OH}
\end{figure}

Now the electronic properties of the ZGNRs are studied, and
asymmetric edge terminations are shown in Figure 1. As revealed in
the previous study\cite{Louie-N}, an important feature of the ground
states of ZGNRs-H-H is that their energy bands are spin degenerate
(shown in Fig. 2(a)). However, it is found that such feature does
not exist in the studied ZGNRs with asymmetric edge terminations,
and their energy gaps for one spin orientation reduce remarkably. In
the following, 8-ZGNRs are chosen as examples for detailed
discussions, because ZGNRs with different width exhibit similar
basic characteristics of electronic structure.

A typical band structure and density of states (DOS) for 8-ZGNR-H-OH
are shown in Figures 2(d, e and f). Compared with 8-ZGNR-H-H, energy
bands for opposite spin are no longer degenerate, and the energy gap
of spin-up bands (0.32 eV) is larger than that of spin-down bands
(0.12 eV). It is found that this spin splitting arises from the
influence of asymmetric edge terminations on the electrostatic
potentials of edge carbon atoms. For 8-ZGNR-H-H, projected DOS
(PDOS) of edge carbon atoms (Fig. 2(b)) indicates that the energies
of edge states localized at left and right edges are the same due to
the equal electrostatic potentials at two edges, so the energy gaps
associated with spin-up and spin-down are equal. Whereas for
8-ZGNR-H-OH, because of the difference of electrostatic potentials
at two edges caused by asymmetric edge terminations, the energies of
edge states localized at left (H-terminated) edge are lower than
that of states localized at right (OH-terminated) edge, as PDOS of
edge carbon atoms (Fig. 2(c)) illustrated. In addition, the states
localized at two edges for each spin are occupied and unoccupied
states respectively, so the energy difference of left and right edge
states leads to the unique behavior that the energy gap associated
with one spin increases, whereas that associated with the other
decreases (as shown in Fig. 2(c)).

\begin{figure}
\includegraphics[width=0.475\textwidth]{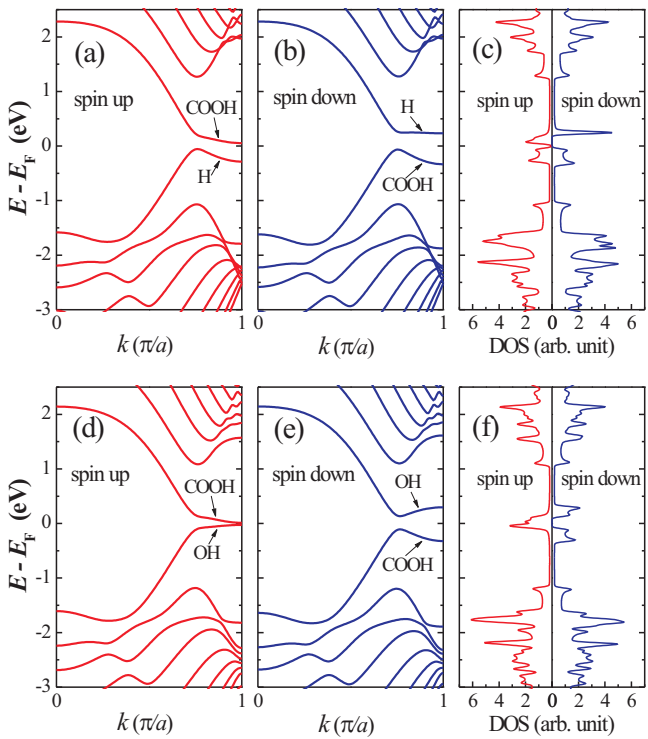}
\caption{(a) Spin-up and (b) spin-down energy bands, and (c) the
total DOS of 8-ZGNR-H-COOH in AF configuration. (d) Spin-up and (e)
spin-down energy bands, and (f) the total DOS of 8-ZGNR-OH-COOH in
AF configuration. In (a), (b), (d) and (e), the arrows indicate the
edge states and which edge they are localized at.}\label{2bands}
\end{figure}

Like 8-ZGNR-H-OH, the energy bands of 8-ZGNR-H-COOH in the
anti-ferromagnetic configuration (shown in Figs. 3(a and b)) are
also spin split around the Fermi level because of the asymmetry of
edge terminations. In contrast to 8-ZGNR-H-OH, however, the relative
small energy gap (0.11 eV) appears in the spin-up bands rather than
in the spin-down bands, and the energies of states localized at
H-terminated edge are higher than that of states localized at
COOH-terminated edge. This difference indicates that the changes of
the electrostatic potential of edge carbon atoms induced by OH and
COOH are opposite, and implies that 8-ZGNR-OH-COOH would have a
larger difference between electrostatic potentials at two edges and
a smaller energy gap for one spin orientation. Our further
calculations confirm this viewpoint and show that the energy gap
associated with spin-down is 0.24 eV, and that associated with
spin-up is only 0.04 eV (Figs. 3(d) and 3(e)). Therefore, the ground
state of 8-ZGNR-OH-COOH is very close to half-metallic state.

\begin{figure}
\includegraphics[width=0.475\textwidth]{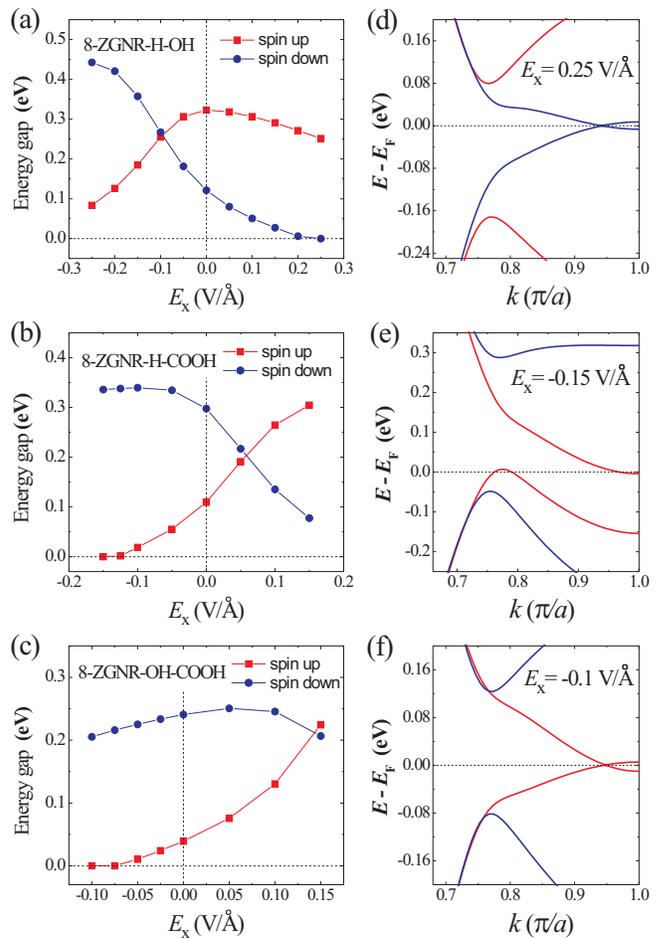}
\caption{Electric field effect on the spin-polarized bandgaps of (a)
8-ZGNR-H-OH, (b) 8-ZGNR-H-COOH, and (c) 8-ZGNR-OH-COOH, where
positive $E_{x}$ represents that the direction of electric fields is
from left edge to right edge of ZGNRs, while negative $E_{x}$
represents the opposite direction. (d), (e), (f) The spin-resolved
band structures corresponding to (a), (b), and (c) when the systems
achieve the half-metallic states, where red (blue) lines represent
spin up (down).}\label{efield}
\end{figure}

As mentioned above, the ZGNRs considered in our study maintain the
edge states and anti-ferromagnetic configuration like ZGNRs-H-H.
Therefore, one would expect that these systems behave as half-metals
under the influence of an appropriate external electric field. To
substantiate this presumption, we have calculated the energy gaps of
spin-up and spin-down bands as a function of the intensity of an
external electric field along the $x$ direction ($E_{x}$), which are
shown in Figure 4, where positive $E_{x}$ represents that the
direction of electric fields is from left edge to right edge of
ZGNRs, while negative $E_{x}$ represents the opposite direction.

For 8-ZGNR-H-OH, Figure 4(a) shows that the energy gap of spin-down
bands decreases successively as $E_{x}$ increases, and becomes zero
when $E_{x}$ reaches 0.21 V/\AA~. At the same time, the energy gap
of spin-up bands remains very large, which means that the system has
become half-metallic. Similar to 8-ZGNR-H-OH, 8-ZGNR-H-COOH and
8-ZGNR-OH-COOH also exhibit half-metallic behaviors under the
influence of an external electric field, as shown in Figures 4(b and
c). However, it is noticeable that for 8-ZGNR-H-COOH and
8-ZGNR-OH-COOH, the $E_{x}$ that make the systems achieve
half-metallic states is negative rather than positive like
8-ZGNR-H-OH, because the energies of left edge states are lower than
that of right edge states in 8-ZGNR-H-OH (Fig. 2(c)), whereas the
situation is opposite in 8-ZGNR-H-COOH and 8-ZGNR-OH-COOH. Thus, the
directions of electric fields that can efficiently force the systems
into half-metallic states are different for these ZGNRs.

It is interesting to note that the absolute values of critical
electric fields required to achieve half-metallic states for the
later two kinds of ZGNRs are 0.13 and 0.07 V/$\rm{\AA}$
respectively, which are much smaller than that of 8-ZGNR-H-H (this
value given by our calculations is 0.22 V/$\rm{\AA}$). Such
characteristic results from the fact that the energy gap for one
spin orientation has already reduced without external electric
fields. Therefore, the half-metallic states can be achieved more
easily due to asymmetric edge terminations, and such ZGNRs could
serve as the spin filter devices even without the application of an
external electric field. Figures 4(d, e and f) show the
spin-resolved band structures of three studied ZGNRs in the
half-metallic state. It is found that for ZGNRs-H-OH and
ZGNRs-OH-COOH, only two bands for one spin orientation cross at the
Fermi level, which indicates that their charge carriers mimic
massless Dirac fermions similar to the experimental observation in
graphene\cite{Novo-N}. However, this property does not exist in
ZGNRs-H-H\cite{Louie-N} and ZGNRs-H-COOH (Fig. 4(e)) since they have
indirect energy gaps before they are forced into half-metallic
states.

\

\noindent \textbf{4. CONCLUSION}

\

\noindent In conclusion, the influence of asymmetric edge
terminations on the spin-polarized electronic structure of ZGNRs is
studied using \emph{ab initio} calculations. It is found that
different edge termination groups can change the energies of edge
states of ZGNRs without breaking the original spin-polarized
anti-ferromagnetic configuration, thus asymmetric edge terminations
can lead to a spin splitting of energy bands around the Fermi level.
Most importantly, the energy gap of one spin orientation almost
disappears, which means the ground state is very close to the
half-metallic state. Thus, changing edge termination groups via
chemical treatments is an effective way to reduce the required
critical electric field to make ZGNRs achieve half-metallic states,
and even can make their ground states half-metallic directly. This
effect provides a promising possibility of using ZGNRs as spin
filter devices in future nanospintronics.

\

\noindent\textbf{Acknowledgments:} This work was supported by the
National Natural Science Foundation of China (Grant Nos. 10325415
and 10674077) and the Ministry of Science and Technology of China.


\begin{references}
\bibitem{BE} The binding energy is defined as
$E$(ZGNR-H-FG)-$E$(ZGNR-H)-$E$(FG), where $E$(ZGNR-H-FG),
$E$(ZGNR-H) and $E$(FG) are the total energies of ZGNR terminated
by H and a functional group (FG) at opposite edges, ZGNR
terminated by H at one edge, and a functional group respectively
in a supercell.
\end{references}
\end{document}